\documentclass{jfm}
\usepackage{graphicx}
\usepackage{epstopdf, epsfig}
\usepackage{xcolor}
\usepackage{amsmath}

\shorttitle{Slip due to Kink Propagation at the Liquid-Solid Interface}
\shortauthor{Metehan Cam, Christopher G. Goedde and Seth Lichter}

\title{Slip due to Kink Propagation at the Liquid-Solid Interface}

\author{Metehan \c{C}am\aff{1},
  Christopher G. Goedde\aff{2}
 \and Seth Lichter\aff{1}  \corresp{\email{s-lichter@northwestern.edu}}}

\affiliation{\aff{1}Department of Mechanical Engineering, Northwestern University,
Evanston, IL 60208, USA
\aff{2}Department of Physics and Astrophysics, DePaul University, Chicago, IL 60614, USA}

\begin{document}

\maketitle

\begin{abstract}
\noindent
In Couette flow, the liquid atoms adjacent to a solid substrate may
have a finite average tangential velocity relative to the substrate.
This so-called slip has been frequently observed.
However, the particular molecular-level
mechanisms that give rise to liquid slip are poorly understood. 
It is often assumed that liquid slip occurs by surface diffusion whereby atoms independently move from one substrate equilibrium site to another. 
We show that under certain conditions, liquid slip is due not to singular independent molecular motion, 
but to localized nonlinear waves that 
propagate at speeds that are orders of magnitude greater than
the slip velocity at the liquid-solid interface.
Using non-equilibrium molecular dynamics simulations, we find the properties of these
waves and the conditions under which they 
are to be expected as the main progenitors of slip.
We also provide a theoretical guide to the properties of these nonlinear waves by using an augmented Frenkel-Kontorova model.
The new understanding provided by our results may lead to enhanced capabilities of the liquid-solid interface, for heat transfer, mixing, and surface-mediated catalysis.
\end{abstract}
\noindent 

\clearpage

\section{Introduction}

Slip velocity can be defined as the average tangential velocity of those liquid atoms adjacent to a solid surface.
This average motion has been frequently observed (\cite{Lauga2007,barrat1999influence}).
However, observations of and knowledge about average motion 
conceal the episodic and short-lived events that constitute slip.
We report on transitory groups of atoms with rapid mobility over the solid surface.
These localized groups can be the predominant source of slip, rendering the contribution from surface diffusion negligible.  
In this paper, we investigate and characterize this unheralded mechanism of slip.

We show that, at any instant, most liquid atoms adjacent to the solid do not contribute to the slip velocity.
Rather, it is only a few groups of liquid atoms that coordinate their motions into fast-moving localized nonlinear waves that convect mass.
Under some conditions, the slip velocity is due nearly entirely to the sum of the displacements accrued by these localized waves. 
In these cases, slip is not due to surface diffusion, but to wave propagation.
In their review of slip, \cite{Lauga2007} emphasize 
the complexity of phenomena that take place at the liquid-solid interface.
With the results presented here, we add another element to their array of interesting 
phenomena that contribute to slip.

There have been many elegant and artful numerical, analytical, and experimental studies of liquid slip, summarized in $\S$\ref{sec.prior_research}, including molecular dynamic simulations similar to our simulations, which are described in $\S$\ref{sec.md_slip}.  
We find a previously unobserved phenomenon of rapidly propagating waves at the liquid-solid interface.
This finding is made possible by the insight that the solitonic nature of localized disturbances
produces a unique signature.
That signature is the doubly-occupied lattice cell, made visible by measuring the locations of the liquid atoms relative to the substrate lattice, as we describe in $\S$\ref{sec.mechanism_slip}.
We can then scan our data for the occurrence of a doubly-occupied cell in order to extract those events most meaningful for slip.
In particular, the doubly-occupied cell is central to a localized nonlinear wave that transports mass and leads to slip, as described in $\S$\ref{sec.FK_kinks}.
Observations from our molecular dynamics (MD) simulations agree well with the predictions of the Frenkel-Kontorova and sine-Gordon equations, lending analytical support characterizing our observations as nonlinear waves akin to solitons, as further discussed in $\S$\ref{sec.FK_kinks}.
We summarize the main points of our study and note their applicability in the final $\S$\ref{sec.conclusions}.

\section{Prior Research} \label{sec.prior_research}

The no-slip boundary condition remains a dependable boundary condition for many macroscopic Newtonian flows. 
However, when liquids are confined to channels of a width of only a few molecular diameters, the conventional hydrodynamic theories based on the Navier-Stokes equations may fail, and the no-slip boundary condition may no longer be valid (\cite{travis1997departure,travis2000poiseuille,zhu2002limits}). Experimental, analytical, and numerical studies showed the now well-known slip: that liquids adjacent to a solid move with a finite velocity relative to the solid surface (\cite{Lauga2007,Shu2018,Wang2021}). 
Recently, \cite{Hadjiconstantinou2024} presented a discussion of the state of the art of the theoretical understanding of slip. 
Here, we further discuss prior works that are most relevant to our study.

Together with the equations that now bear his name,
Navier derived what became known as the Navier slip condition (\cite{Navier1822}),
\begin{equation}
E U + \epsilon\frac{dU}{dz} = 0,
\end{equation}
shown here for the tangential component of the liquid velocity $U$ above a flat wall whose normal is in the $z$-direction.
Navier delegated constants $E$ and $\epsilon$ as sliding resistances; $\epsilon$ as the sliding resistance between fluid molecules in adjacent layers, the present-day dynamic viscosity, 
and $E$ as the sliding resistance between the molecules in the vicinal liquid layer and the wall itself, the present-day liquid-solid friction coefficient.
The ratio $\epsilon/E$ is the slip length. Navier's presentation was in terms of molecules moving in layers, a point of view that we also use in defining the first liquid layer (FLL) in $\S$\ref{sec.md_slip}.
In fact, near a wall, the liquid density is layered (\cite{Rowley1976,Abraham1978}).
Careful formulations and analyses of the slip boundary condition can be found in
\cite{Miksis1994},
\cite{Brenner2000}, and 
\cite{Bolanos2017}, including over textured and patterned surfaces 
(\cite {Priezjev2006,Kamrin2010,Luchini2013,Zampogna2019}).

\cite{Koplik1989} provided an early yet thorough MD simulation of liquid slip including examination of Couette and Poiseuille flows, contact line motion, and the trajectories of the individual liquid atoms. Our MD simulation methodology is similar to theirs in that we make use of molecular trajectories to identify the atomic-level mechanisms of liquid slip, as described in $\S$\ref{sec.mechanism_slip}. 

Models of surface diffusion at the liquid-solid interface have presented the dependence of slip on shear rate, temperature, and molecular interaction parameters (\cite{Hsu2010,Shu2018,Wang2019,Wang2021,Shan2022}).
\cite{Hadjiconstantinou2021} formulated and solved 
the Fokker-Planck  
dynamics of a point particle 
moving in one dimension, driven by shear, 
slowed by friction with a periodic substrate,
and interacting with a thermal background.
As shear force is increased, these results reveal
slip length at first insensitive to shear, then rapidly increasing
to a higher value that once again is invariant to shear.
As pointed out by Hadjiconstantinou, these general 
features of the slip length are 
consistent with those that were found 
by other means (\cite{Martini2008b}).

Motions more complex than simple diffusion have been anticipated 
to contribute to the surface transport at the liquid-solid interface.
In particular, several types of mechanisms have been described, 
whereby atoms' correlated motion with neighboring atoms has been described 
as gliding, reptation, and dislocation motions (\cite{Oura2013}).
Studies of monolayers range from numerical studies of atoms on atomically-structured substrates (\cite{Cam2021}) to micron-sized beads over a laser-generated interference pattern
(\cite{Reguzzoni2010,Brazda2018}).
These studies reveal that even at low levels of forcing, patches
of molecules will slip
due to the incommensurate
alignment between the particles
and the surface.
In contrast, observations of 
coordinated motion at the liquid-solid interface under 
macroscopic flows such as Couette or Poiseuille flow are lacking.
It is precisely to seek such coordinated motion that we 
undertook this investigation.

\newpage

\section{MD Simulations of Liquid Slip} \label{sec.md_slip}

\begin{figure}
  \hspace*{-0.2in}
  \centering
  \includegraphics[scale=0.40]{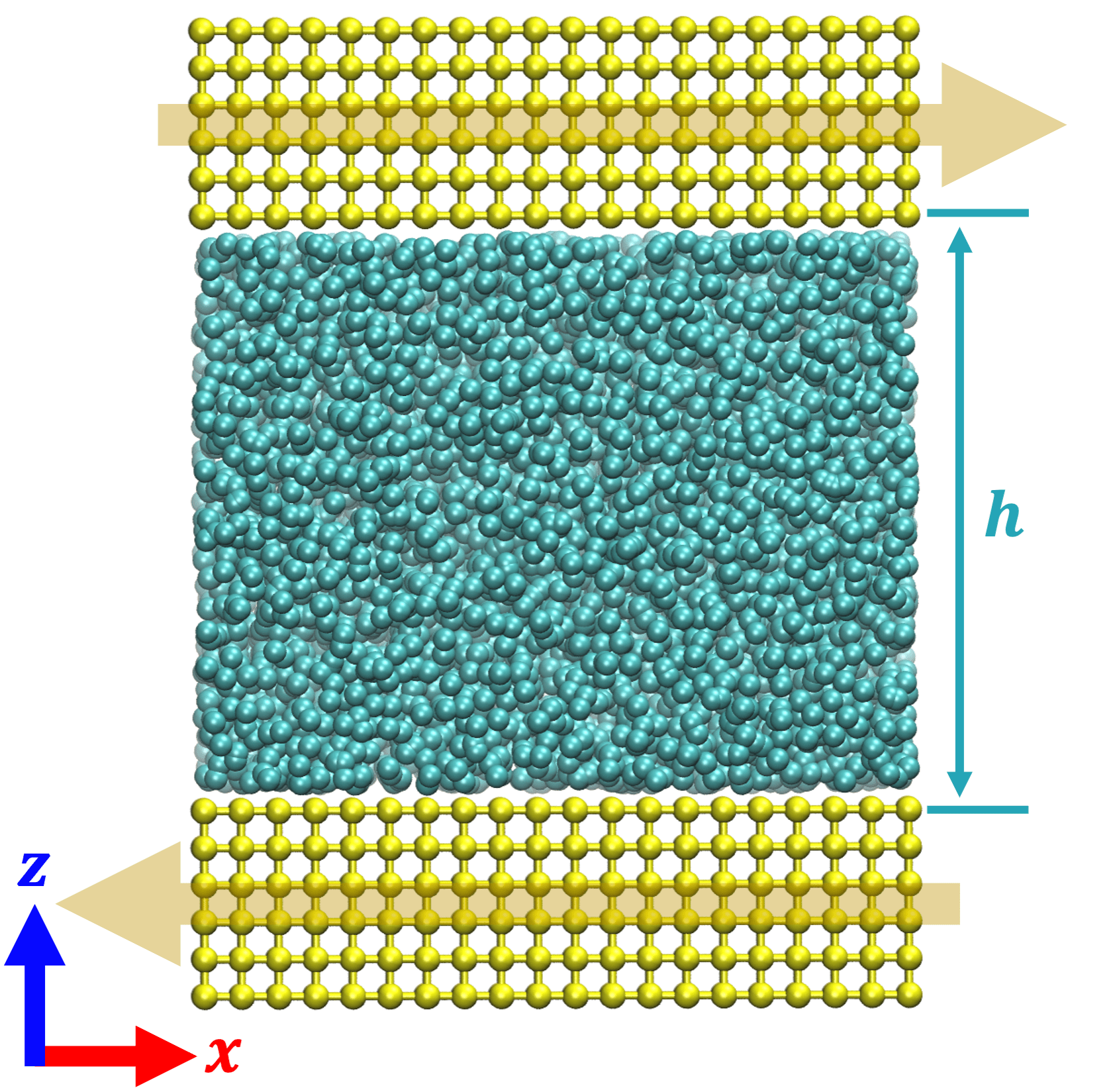}
  \centering
   \caption{Snapshot of the Couette flow setup. 
   Liquid and solid atoms are shown as cyan and yellow spheres, respectively.
    Solid yellow lines indicate the linear bonds between the nearest-neighbor solid atoms. 
   The direction of motion of each solid wall is shown
   by a large yellow arrow. 
   The channel width $h$ is the $z$-distance 
   between the mean positions of the inner-most solid layers.}
  \label{fig:Couette_Setup}
\end{figure}

Our nonequilibrium molecular dynamics simulations are performed using the Nanoscale Molecular Dynamics (NAMD) program (\cite{phillips2005scalable}). We simulate a liquid confined between two thermal solid walls, shown in figure~\ref{fig:Couette_Setup}. 
Each wall consists of six 20 by 20 layers of atoms in the $x$- and $y$-directions.
The solid atoms are maintained 
on a cubic lattice
with a lattice spacing of $\lambda$ by nearest-neighbor linear springs of stiffness 20~N/m.
Periodic boundary conditions are implemented along the $x$- and $y$-directions of the square domain, which has dimensions of $L_{x} = L_{y} = 20\lambda$. Periodic boundary conditions are also used in the $z$-direction, with the periodic box size chosen to be large enough that the periodic images of the upper and lower walls are always separated by a distance greater than the NAMD non-bonded interaction cutoff distance of 1.4 nm.
The channel width, $h$, equal to the $z$-distance between the mean positions of the inner-most solid layers, is eighteen liquid molecular diameters. 

The non-bonded interaction potentials are given by
\begin{equation}
    \displaystyle \phi_{ij}(r) = 4\epsilon_{ij} \left[ \left( \sigma_{ij}/r \right)^{12} - \left( \sigma_{ij}/r \right)^{6} \right],
    \label{eq:LJ}
\end{equation}
where $\phi_{ij}$ indicates the Lennard-Jones (LJ) interaction strength between atoms of types $i$ and $j$. 
Indices $i$ and $j$ are chosen to indicate the atomic species, liquid (L) or solid (S). 
The Lennard-Jones potential provides a simple model of atomic interaction, with only an energy parameter $\epsilon_{ij}$ and a length parameter $\sigma_{ij}$.   
The energy and length parameters associated with the interactions between the liquid and solid atoms are determined using the Lorentz-Berthelot combining rules (\cite{allen2004introduction}),
\begin{equation}
    \displaystyle \sigma_{\mathrm{LS}} = \frac{\sigma_{\mathrm{LL}} + \sigma_{\mathrm{SS}}}{2} \quad \mathrm{and} \quad \epsilon_{\mathrm{LS}} = \sqrt{\epsilon_{\mathrm{LL}}\,\epsilon_{\mathrm{SS}}} \, .
    \label{eq:LB}
\end{equation}
Table~\ref{tab:LJ_Parameters} summarizes the energy and length parameters and the masses of the liquid and solid atoms. In all simulations, the length parameter for the solid-solid interaction is chosen such that the non-bonded equilibrium spacing of the solid atoms, $r_{0}^{\mathrm{SS}}\equiv 2^{1/6}\sigma_{\mathrm{SS}}$, is equal to the solid lattice spacing $\lambda$.
\begin{table}
  \begin{center}
\def~{\hphantom{0}}
  \begin{tabular}{lccc}
      $\mathrm{Atom \, Type}$  & $\epsilon_{ij}/k_{\mathrm{B}}$ (K)   &   $\sigma_{ij}$ (nm) & $m_{i}$ (amu) \\[3pt]
       Liquid (LL)   & 188 & 0.266 & 14\\
       Solid (SS)   & 47 & 0.119-0.341& 197\\
       Liquid-Solid (LS)  & 94 & 0.193-0.303& -\\
  \end{tabular}  \caption{\label{tab:LJ_Parameters} Lennard-Jones energy $\epsilon_{ij}$ and length $\sigma_{ij}$ parameters associated with the liquid-liquid (LL), solid-solid (SS), and liquid-solid (LS) interactions as well as the masses of liquid and solid atoms. $k_\mathrm{B}$ is the Boltzmann constant.}
  \label{tab:kd}
  \end{center}
\end{table}

In some MD simulations, the solid atoms bounding the liquid flow are fixed in their lattice positions and not given the freedom to fluctuate, constituting a so-called rigid or athermal wall.
\cite{Martini2008a} have shown that at high shear rates, slip is more accurately reproduced by allowing the wall atoms to possess their thermal energy and fluctuate, as we do in this study.
Furthermore, liquid atomic trajectories are most faithfully simulated when the viscous heat generated by the sheared liquid is allowed to flow to the solid boundaries and, only there, be thermostated (\cite{bernardi2010thermostating,yong2013thermostats}).
Therefore, in this study, we thermostat neither the liquid itself nor the adjacent solid layers, but thermostat only the outer solid layers most distant from the liquid. 

The simulation protocol consists of three stages. 
In the first stage, the walls are not yet set into motion while the entire system is coupled to the Langevin thermostat.
This first stage allows the liquid atoms to diffuse into random positions.
In the second stage, the thermostat is restricted only to the two outer solid layers in each wall. 
During this stage, the top and bottom solid walls accelerate to their target velocities,  $U_{\mathrm{WALL}}$, in the $x$-direction, as shown by the horizontal yellow arrows in figure~\ref{fig:Couette_Setup}. 
Target velocities of the top and bottom solid walls are equal in magnitude but opposite in direction. 
In the third stage, both solid walls move with constant velocity in the $x$-direction. 
We report only the results obtained during the third stage, in which the average properties of the liquid and solid remain constant. 

As discussed in detail below, we carry out two parametric studies.
In one, the wall velocity, $U_\mathrm{WALL}$, is systematically varied from 0.2 to 220 m/s. 
For these studies we present the slip velocity as a function of shear rate. 
We calculate the shear rate by finding the slope of a straight-line fit to the liquid velocity profile in $z$-direction near the center of the channel. 

Secondly, we present results where the Lennard-Jones size of the liquid atoms $r_{0}^{\mathrm{LL}}\equiv 2^{1/6}\sigma_{\mathrm{LL}}$ is kept constant while the lattice spacing $\lambda$ and the solid-solid length parameter $\sigma_\mathrm{SS}$ are varied, yielding the ratio $r_{0}^{\mathrm{LL}}/\lambda$ of the relative size of the liquid atoms to the solid atoms as the control parameter.

\begin{figure}
  \hspace*{-0.2in}
  \centering
  \includegraphics[scale=0.50]{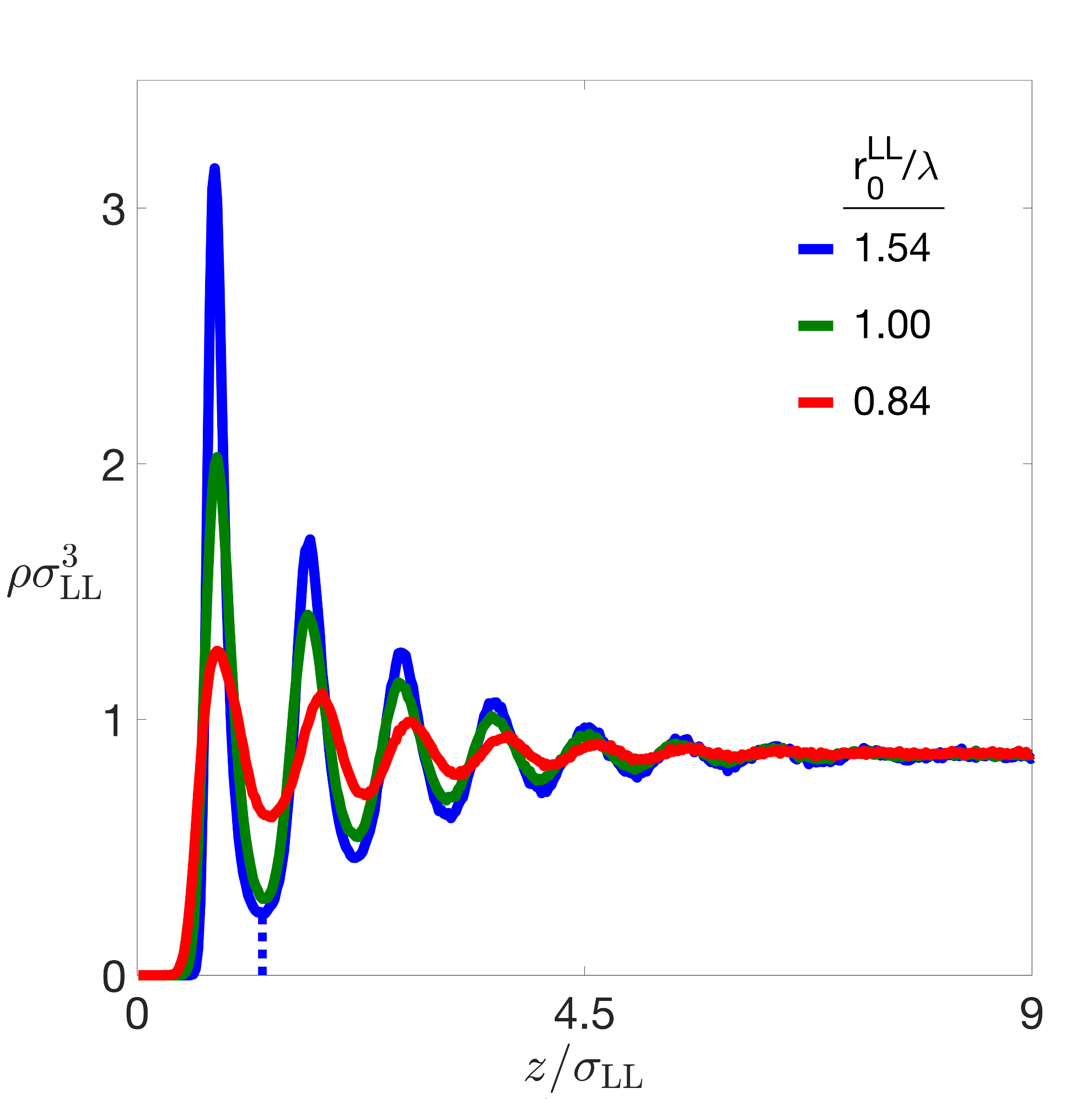}
  \centering
   \caption{The liquid density profiles, $\rho\sigma_{\mathrm{LL}}^{3}$, as a function of the $z$-position of the liquid atoms, $z/\sigma_{\mathrm{LL}}$. $z$ is measured relative to the inner-most solid layer. Blue, green, and red solid lines indicate the liquid density profiles of $r_{0}^{\mathrm{LL}}/\lambda=$ 1.54, 1.00, and 0.84, respectively.  
   The first minimum defining the upper boundary of the first liquid layer is nearly the same for all densities, and is shown by a dotted blue line for 
   $r_{0}^\mathrm{LL}/\lambda=1.54$.
   Far enough from the solid walls, the liquid densities asymptote to a common value of  $\rho\sigma_{\mathrm{LL}}^{3}=0.87$.}\label{fig:Density_Profiles}
\end{figure}

For data collection at a sampling rate of 1~ps, the duration of the third stage is 4 to 26 ns, depending on the shear rate.
For each value of  $r_{0}^{\mathrm{LL}}/\lambda$, we run four or five simulations with distinct initial positions and velocities for the liquid atoms, while all other simulation parameters remained the same.
Sampling at 1~ps would be sufficient if our only objective is to measure the average properties of the system such as temperature, density, slip velocity, and slip length. 
However, it typically takes less than a picosecond for a liquid atom to move from one substrate equilibrium site to the next. 
Consequently, for our goal to resolve the detailed trajectories of the coordinated motion of individual atoms, femtosecond resolution in atomic positions is needed. 
In simulations where we gather data on these dynamics, we use a sampling rate of 10~fs and run the simulations up to 2~ns.

A well-established observation of atomic-level structure is the variation in liquid density near a solid (\cite{Rowley1976,Abraham1978,barrat1999influence,morciano2017nonequilibrium}). 
We too find this layering and make use of it in our analysis.
In particular, we define
the first liquid layer as the volume
between the mean positions of the innermost solid layer and the first minima of the liquid density profile, as shown in figure~\ref{fig:Density_Profiles}.
Liquid atoms whose position lie within this volume constitute the first liquid layer.
Far enough from the solid walls, the variations disappear and the liquid density becomes constant.
The total number of liquid atoms in the channel is chosen such that the liquid density, denoted by $\rho$, has the constant value 
$\rho=0.87\sigma_{\mathrm{LL}}^{-3}$ at the center of the channel. 

\begin{figure}
  \hspace*{-0.2in}
  \centering
  \includegraphics[scale=0.50]{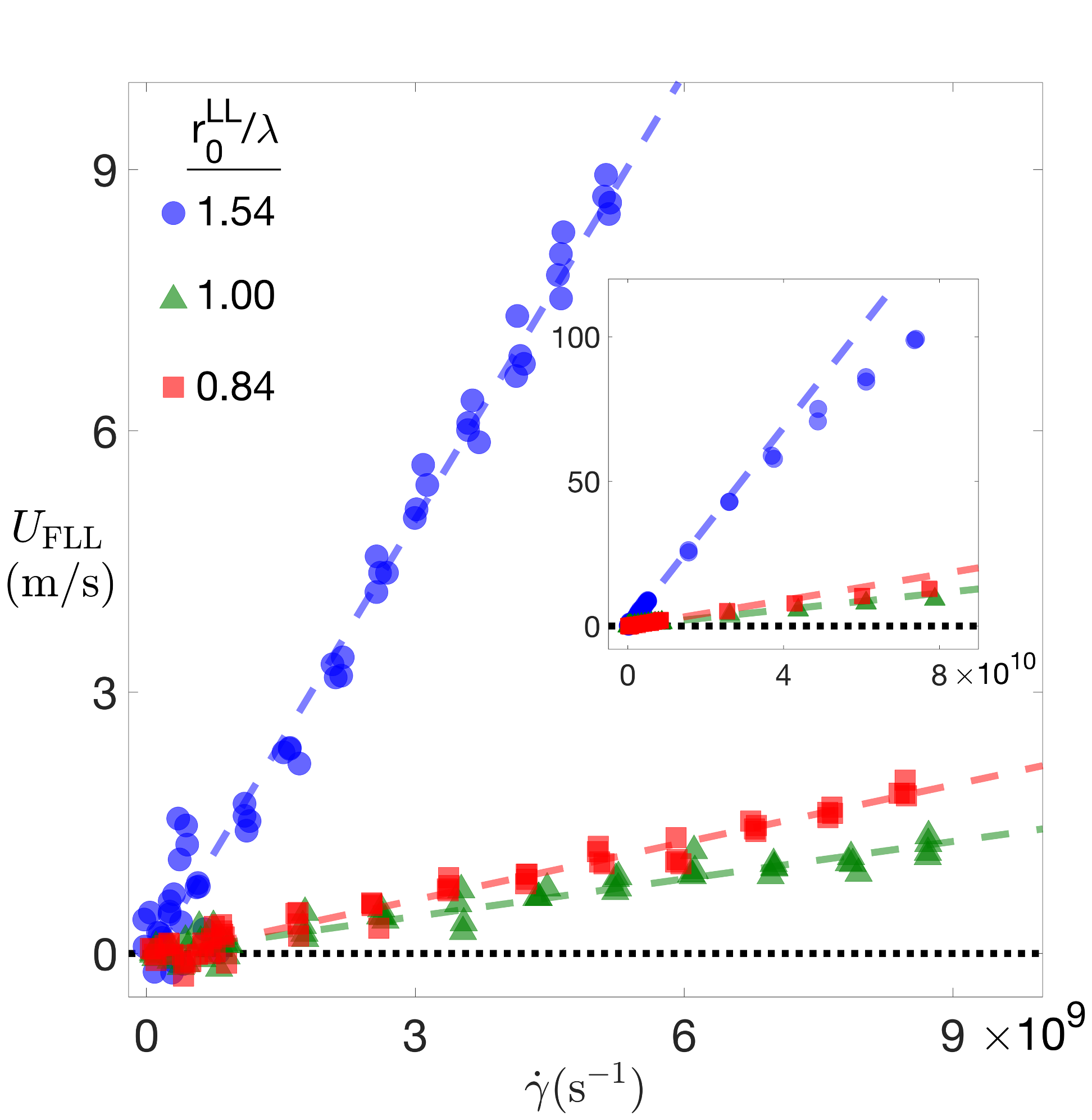}
  \centering
   \caption{The slip velocity of the first liquid layer, $U_{\mathrm{FLL}}$, as a function of the shear rate, $\dot{\gamma}$. Blue circles, green triangles, and red squares indicate the slip velocities as
   the substrate wavelength $\lambda$ increases relative to the size 
   $r_0^{\mathrm{LL}}$ of the liquid atoms,
      $r_{0}^{\mathrm{LL}}/\lambda=$ 1.54, 1.00, and 0.84, respectively.  The dashed lines are linear fits to $U_{\mathrm{FLL}}$ for shear rates $\dot{\gamma}<10^{10} \, \mathrm{s^{-1}}$. 
   The slopes of these lines yield Navier slip lengths of 
   (6.47, 0.54, 0.84)$\,\sigma_{\mathrm{LL}}$ for blue, green, and red, respectively.
   The inset shows $U_{\mathrm{FLL}}$ over a larger range of shear rates.}\label{fig:Slip_Velocity_Shear_Rate}
\end{figure}

Averaging the velocities relative to the moving substrate of all first liquid layer atoms in the $x$-direction over a long enough interval of time yields the slip velocity, $U_{\mathrm{FLL}}$, shown in figure~\ref{fig:Slip_Velocity_Shear_Rate}. 
At lower shear rates, longer simulations are required to resolve the ever-decreasing slip velocity.
Simulations at zero shear rate yield zero slip velocity of approximately ±1 m/s. Thus from zero shear rate to all but the largest shear rates shown in the
inset of figure \ref{fig:Slip_Velocity_Shear_Rate}, our data is consistent with the proportionality of shear rate and slip velocity.
The slopes of the straight-line fit to the data shown in figure~\ref{fig:Slip_Velocity_Shear_Rate} yield the Navier slip lengths of (6.47, 0.54, 0.84)$\,\sigma_{\mathrm{LL}}$ for the three values of $r_0^\text{LL}/\lambda$, (1.54, 1.0, 0.84), respectively.

As also shown in the figure, the slip length depends on the lattice spacing of the substrate, with the slip length increasing as the lattice spacing of the substrate decreases. 
Other researchers have shown the dependence of slip velocity (and slip length) on the geometrical and chemical properties of the substrate (\cite{thompson1997general,barrat1999influence}).
In general, then, we find that slip velocity is not a universal value independent of the spacing of the atoms along the surface, but is dependent on the atomic type and the wavelength of the solid substrate. 

\begin{figure}
  \hspace*{-0.2in}
  \centering
  \includegraphics[scale=0.50]{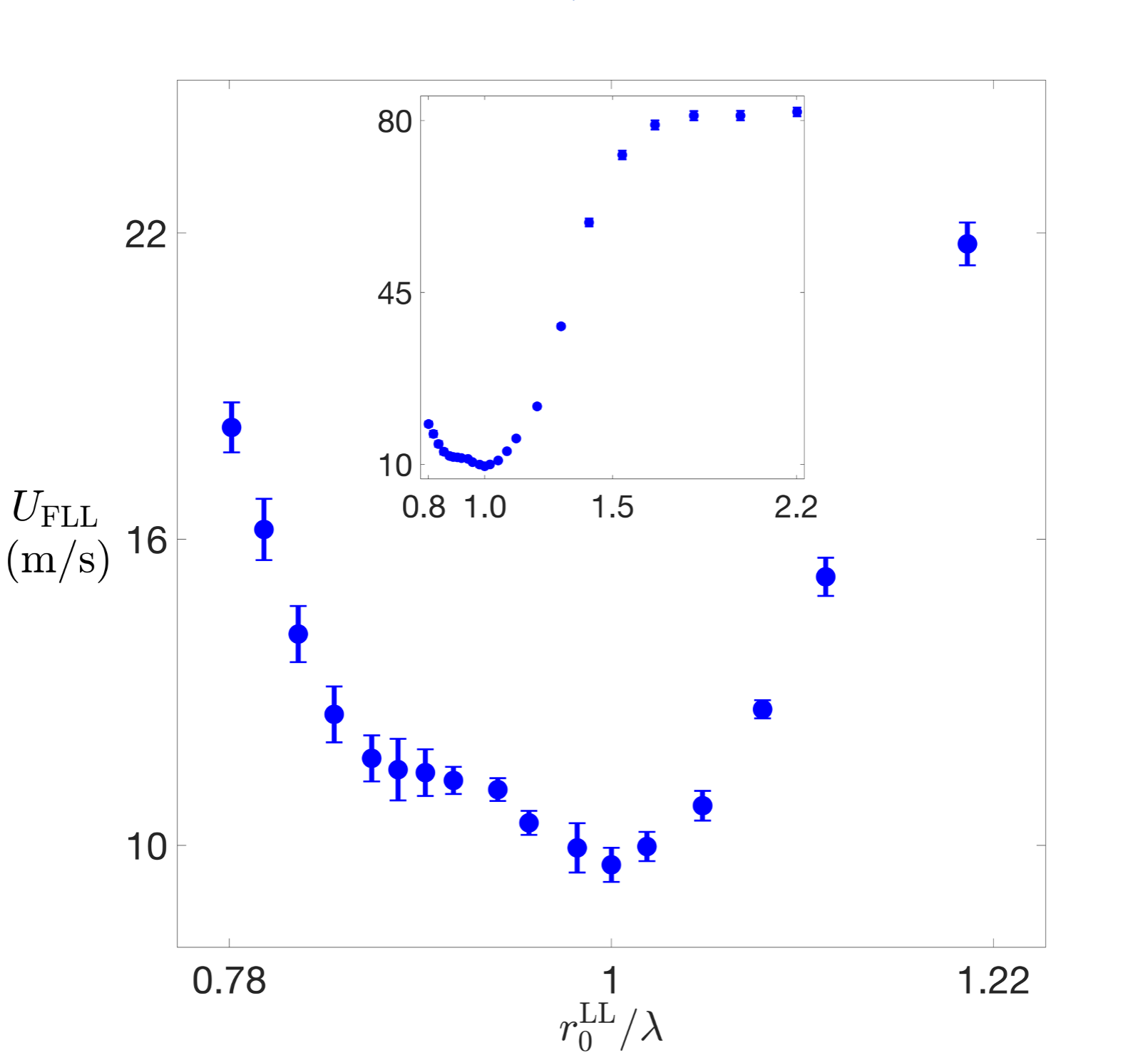}
  \centering
   \caption{A close-up of the slip velocity $U_{\mathrm{FLL}}$ of the first liquid layer for $r_{0}^{\mathrm{LL}}/\lambda$ near unity. 
   The slip velocity has a minimum at $r_{0}^{\mathrm{LL}}/\lambda=1$. 
    Error bars are the standard deviations of five simulations for each $r_{0}^{\mathrm{LL}}/\lambda$. 
   Atomic positions and velocities are sampled at every $1$ ps.
   Wall speed, $U_{\mathrm{WALL}}=180$ m/s. The inset shows $U_{\mathrm{FLL}}$ across 
   a wider range, $0.78 \leq r_{0}^{\mathrm{LL}}/\lambda \leq 2.22$.}\label{fig:True_Slip_Velocity_Low_Shear_Rate}
\end{figure}

Figure \ref{fig:True_Slip_Velocity_Low_Shear_Rate}
shows the dependence of slip velocity on
$r_0^\mathrm{LL}/\lambda$ for a wall speed of $U_\text{WALL}=180$ m/s. 
Ratios $r_0^\mathrm{LL}/\lambda<1$ indicate that the substrate lattice spacing is greater than the equilibrium spacing of the liquid atoms. Conversely, ratios $r_0^\mathrm{LL}/\lambda>1$ indicate that the equilibrium spacing of the solid atoms is smaller than the equilibrium spacing of the liquid atoms. Each data point represents the average of five simulations of duration 6 ns; the error bars represent the standard deviations of each set of five simulations. A minimum in the slip velocity occurs at $r_{0}^\mathrm{LL}/\lambda=1$, where the substrate wavelength is equal to the equilibrium spacing of the liquid atoms. 

\section{Mechanisms of Liquid Slip} \label{sec.mechanism_slip}

\begin{figure}
  \hspace*{-0.0in}
  \centering
  \includegraphics[scale=0.40]{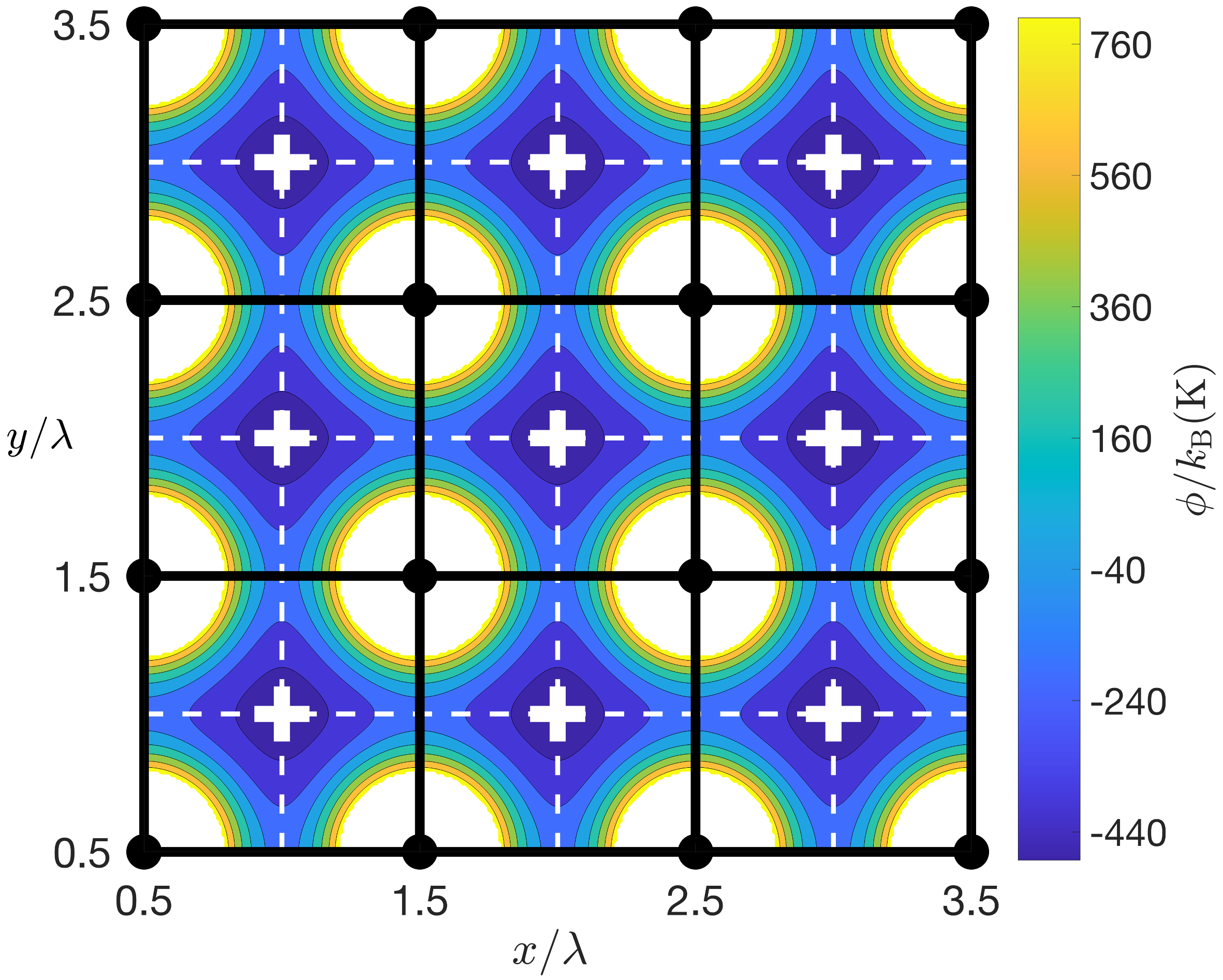}
  \centering
   \caption{Potential energy landscape as a function of the $x$- and $y$-positions of a liquid atom at a fixed height above the substrate. The strength of the potential energy is color-coded: dark blue for low and bright yellow for high potential energy locations. White cross markers indicate the substrate equilibrium sites. 
   Horizontal and vertical white dashed lines indicate low-energy corridors connecting neighboring equilibrium sites. 
   Filled black circles indicate the lattice positions of the solid substrate atoms. 
   Solid vertical and horizontal black lines indicate the boundaries of substrate cells, as defined in the text. A patch of only 9 cells of 400 is shown here.}\label{fig:Cells_Definition}
\end{figure}

The simple cubic arrangement of the substrate atoms creates a square-symmetric potential energy landscape through which the liquid atoms move, as illustrated in figure~\ref{fig:Cells_Definition}. 
On a plane at a small height above the substrate, the largest values of the potential energy are found directly above the substrate atoms.
Conversely, the minimum values of the potential energy are located equidistant from four neighboring solid atoms. We denote these locations, shown as white crosses in the figure, as \textit{substrate equilibrium sites}.
Due to this pattern in the potential energy landscape, the atoms in the first liquid layer are most likely to be found near the substrate equilibrium sites, and they are least likely to be found directly over the locations of the solid atoms. 
As a result, it is unusual to find more than one liquid atom crowded into the neighborhood of an equilibrium site.
Indeed, a doubly-occupied neighborhood 
of an equilibrium site is the critical event for localized slip, as we will show.

Using the pattern shown in figure~\ref{fig:Cells_Definition}, we divide the FLL into 400 cuboidal substrate cells, each centered over a substrate equilibrium site. 
The base of each cell is bounded by four substrate atoms (the black circles in figure~\ref{fig:Cells_Definition})
and the (black) lines connecting them. 
The height of the cells is equal to the height of the first liquid layer.
In figure~\ref{fig:Cells_Definition}, we show 
a top view of 
a patch of 9 neighboring cells (out of 400 total for our 20 by 20 periodic lattice) superimposed on the substrate potential energy landscape.

\begin{figure}
  \hspace*{-0.2in}
  \centering
  \includegraphics[scale=0.35]{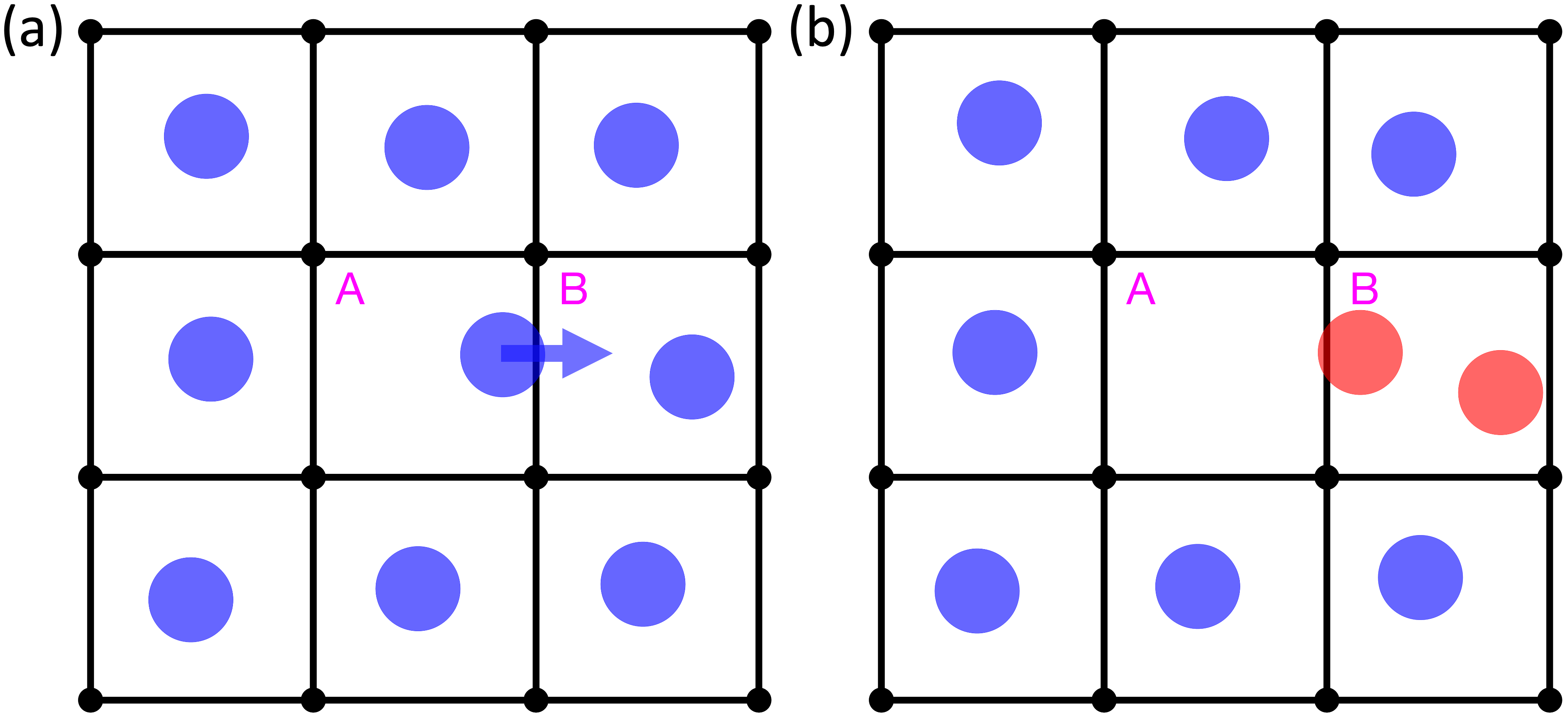}
  \centering
   \caption{Schematic representation of liquid atoms (blue circles) on a small square grid of cells at two consecutive instants of time, shown in panels (a) and (b) respectively.
   Due to the $x$-motion of a liquid atom, shown by the horizontal blue arrow from cells A to B, cell A becomes vacant and cell B becomes doubly-occupied.
   In panel (a), each cell is occupied by a single liquid atom. 
   In panel (b), cell A is not occupied by any liquid atom and is a vacant cell.  
   Cell B, occupied by two liquid atoms (each colored in red), is designated as a doubly-occupied cell. 
    (The diameter of the liquid atoms is chosen for clarity and does not represent their characteristic size.) }\label{fig:Kink_Vacancy_Creation}
\end{figure}

Using the data from simulations sampled every 10 fs, we label and identify all the liquid atoms in the first liquid layer, and track the position of each uniquely labeled atom in time, paying particular attention to the cell that each liquid atom occupies. This allows us to follow each atom in the first liquid layer as it moves between substrate cells. Figure~\ref{fig:Kink_Vacancy_Creation} shows a schematic view of a typical event as an atom in the first liquid layer hops from one cell to the next. In figure~\ref{fig:Kink_Vacancy_Creation}(a), each cell is occupied by a single liquid atom, which is a common configuration of atomic positions in the first liquid layer.
However, a liquid atom can spontaneously move from its current cell into an already-occupied cell, creating a doubly-occupied cell. Figure~\ref{fig:Kink_Vacancy_Creation} shows an example in which a liquid atom moves from the cell labeled A to the cell labeled B in the time between panels (a) and (b), creating doubly-occupied cell B, and leaving behind vacant cell A. 

\begin{figure}
  \hspace*{-0.2in}
  \centering
  \includegraphics[scale=0.35]{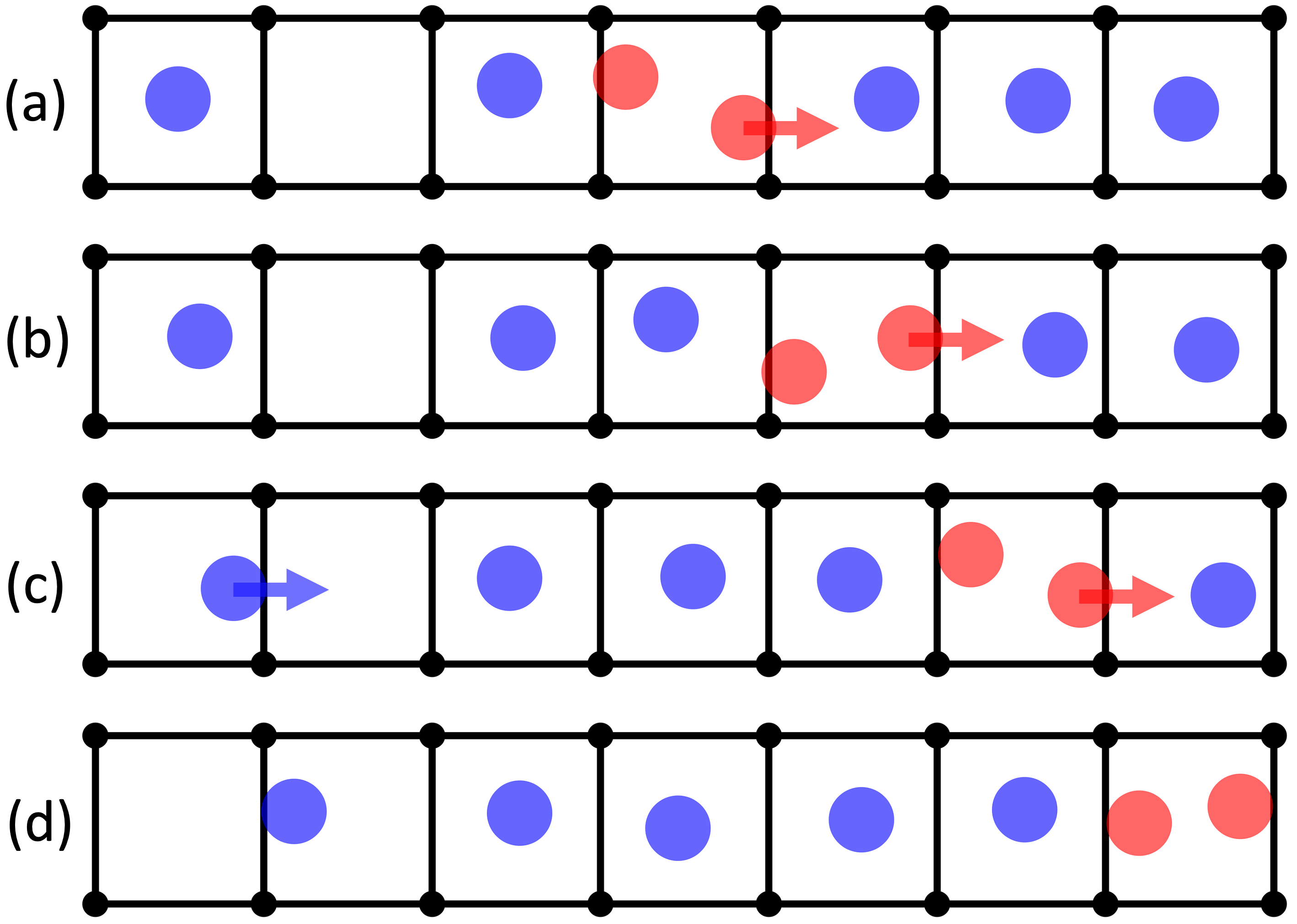}
  \centering
   \caption{Schematic representation of a doubly-occupied cell moving in the $x$-direction. A horizontal strip of seven neighboring cells in the $x$-direction is shown, sampled at four consecutive instants, from panels (a) to (d). As the doubly-occupied cell moves, atoms advance by one cell in the $x$-direction, and the pair of atoms in the doubly-occupied cell (colored in red) evolves sequentially. Horizontal red arrows indicate the atoms that advance into the next cell. Also shown by the horizontal blue arrow, in the time between panels (c) to (d), is the motion of a liquid atom from a singly-occupied cell into a vacant cell, resulting in the vacant cell moving one cell to the left.}   \label{fig:Kink_Propagation_Schematic_Arrows}
\end{figure}

After an atom moves into an already-occupied cell, the original occupant of the now doubly-occupied cell can become dislodged and move forward, and so the double occupancy moves onward into the adjacent cell.
This process can repeat itself, 
as schematically shown in figure~\ref{fig:Kink_Propagation_Schematic_Arrows}.
As the doubly-occupied cell propagates, atoms advance by one cell forward. 
Therefore, the mean motion of the entire layer can be produced by the few atoms moving quickly and sequentially from one doubly-occupied cell to the next in a coordinated fashion. 

The motion of a liquid atom into a vacant (unoccupied) cell is quite different from the motion of an atom into an already-occupied cell. 
When a liquid atom moves into a vacant cell, nearby liquid atoms are nearly unaffected. 
Therefore, such motion represents an uncoordinated,  independent atomic motion over the substrate, which is the typical description of surface diffusion as a mechanism of slip.

We identify and distinguish every instance of either an atomic hop into an already-occupied cell or into a vacant cell. 
We track the motion in both the $x$- and $y$-directions. 
However, as there is no net motion in the transverse 
$y$-direction, we restrict our analysis here to the $x$-direction only.

Using the data from simulations sampled at 10 fs, we separately calculate the net liquid displacement in the $x$-direction due to (i) atomic motion involving doubly-occupied cells and (ii) atomic motion into vacant cells. When divided by the duration of the simulation and the time-averaged number of atoms in the first liquid layer, these displacements provide the fractional contributions to slip due to atomic motion into doubly-occupied or vacant cells.

\begin{figure}
  \centering
  \includegraphics[scale=0.45]{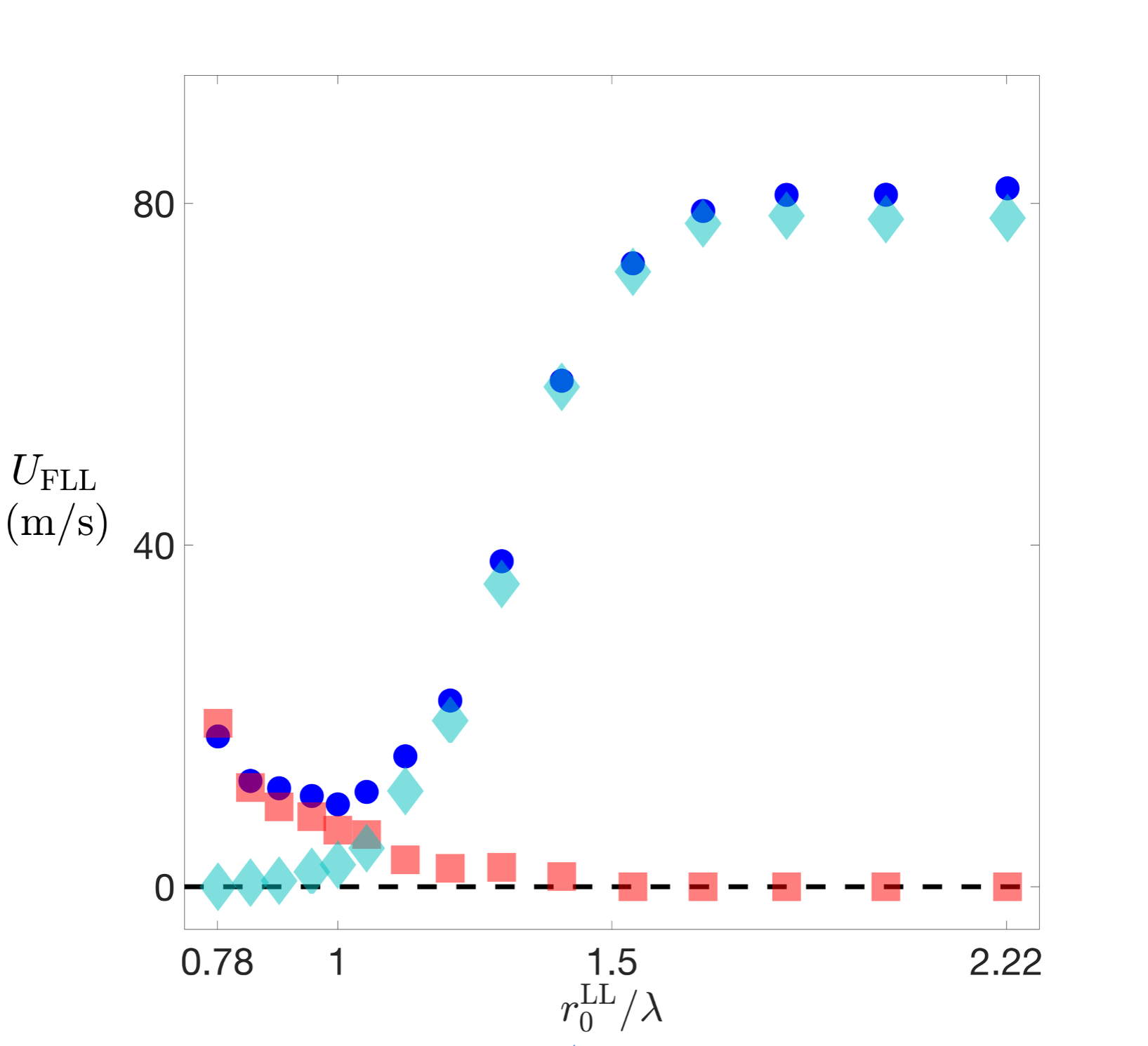}
  \centering
  \caption{The slip velocity of the first liquid layer, $U_{\mathrm{FLL}}$, as a function of $r_{0}^{\mathrm{LL}}/\lambda$, at the wall speed of $U_{\mathrm{WALL}}=180$ m/s. Blue circles indicate the slip velocity as determined by the direct measurement of the atomic velocities. Red squares indicate the slip velocity due to the mass propagation involving doubly-occupied cells alone. Cyan diamonds indicate the slip velocity due to the mass propagation by atoms hopping into vacant cells.
  Atomic positions and velocities are sampled every $10$ fs. }
\label{fig:Slip_Velocity_Kink_Vacancy}
\end{figure}

In figure~\ref{fig:Slip_Velocity_Kink_Vacancy}, we show the slip velocity of the first liquid layer over the range $0.78 \leq r_0^\mathrm{LL}/\lambda \leq 2.22$ for a wall speed of $U_\mathrm{WALL}=180\ \mathrm{m}/\mathrm{s}$. 
For each value of $r_0^\mathrm{LL}/\lambda$, we run a single simulation with a sampling rate of 10 fs and a simulation duration of 2 ns.  
The blue circles show the slip velocity for the layer as calculated directly from the atomic velocities. (This is the same range of data as shown in figure \ref{fig:True_Slip_Velocity_Low_Shear_Rate}.)
The red squares show the slip velocity due only to motion involving doubly-occupied cells. 
The cyan diamonds show only the slip due to atomic motion into vacant cells. 
Taken together, the sum of these two contributions accounts for the totality of the slip. 
At $r_{0}^{\mathrm{LL}}/\lambda>1$, liquid slip predominantly occurs due to mass propagation by atoms hopping into vacant cells. 
Conversely, for $r_{0}^{\mathrm{LL}}/\lambda<1$, liquid slip predominantly occurs due to mass propagation involving doubly-occupied cells. 

\begin{figure}
  \hspace*{-0.3in}
  \centering
  \includegraphics[scale=0.50]{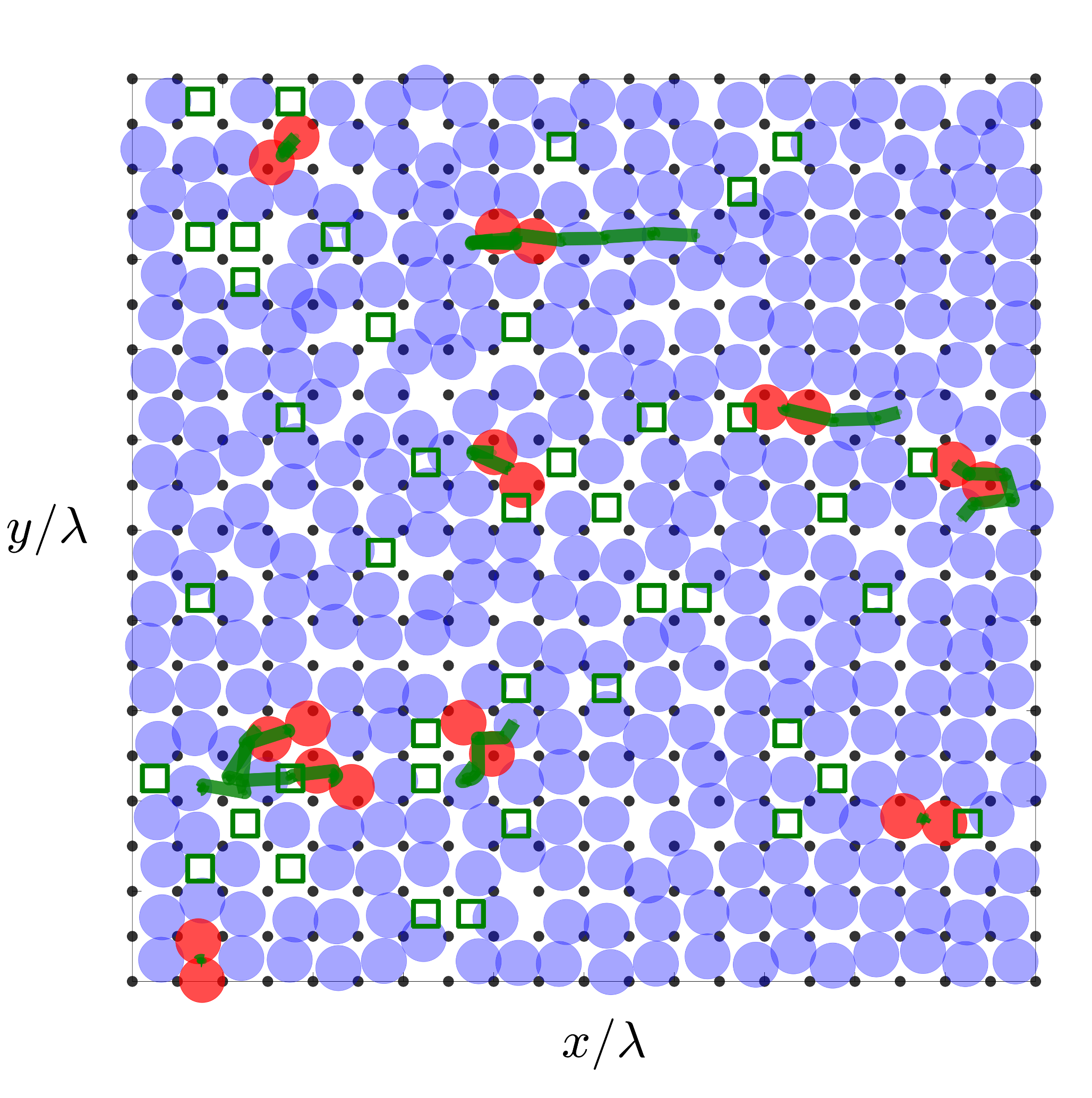}
  \centering
   \caption{Portrait of the first liquid layer constructed from the MD data.  The blue translucent circles indicate the instantaneous positions of the liquid atoms. 
   Liquid atoms in doubly-occupied cells are colored red. 
   The thick green lines indicate the trajectories over time of doubly-occupied cells. 
   The open green squares indicate vacant cells. 
   The black dots indicate the lattice positions of the solid substrate atoms. 
   The mean slip direction is to the right.
   Here, the wall speed is $U_{\mathrm{WALL}}=20$ m/s, $r_{0}^{\mathrm{LL}}/\lambda=1$, and the atomic positions are sampled at every 10 fs.}
\label{fig:Surface_Vis}
\end{figure}

To provide a sense of the role played by vacant and doubly-occupied cells in the generation of slip, figure~\ref{fig:Surface_Vis} shows an annotated view of the entire first liquid layer obtained from the simulation with $r_{0}^{\mathrm{LL}}/\lambda=1$ at the wall speed of $U_{\mathrm{WALL}}=20$ m/s, and a sampling rate of 10 fs. 
Approximately 5\% of the liquid atoms are in a doubly-occupied cell, while approximately 95\% of the atoms merely fluctuate around their equilibrium sites. 
Also shown are the trajectories of doubly-occupied cells as they were tracked moving over several cells; for these parameter values the lifetime of the motion involving a doubly-occupied cell is typically a few picoseconds.

\begin{figure}  \hspace*{-0.25in}
  \centering
  \includegraphics[scale=0.40]{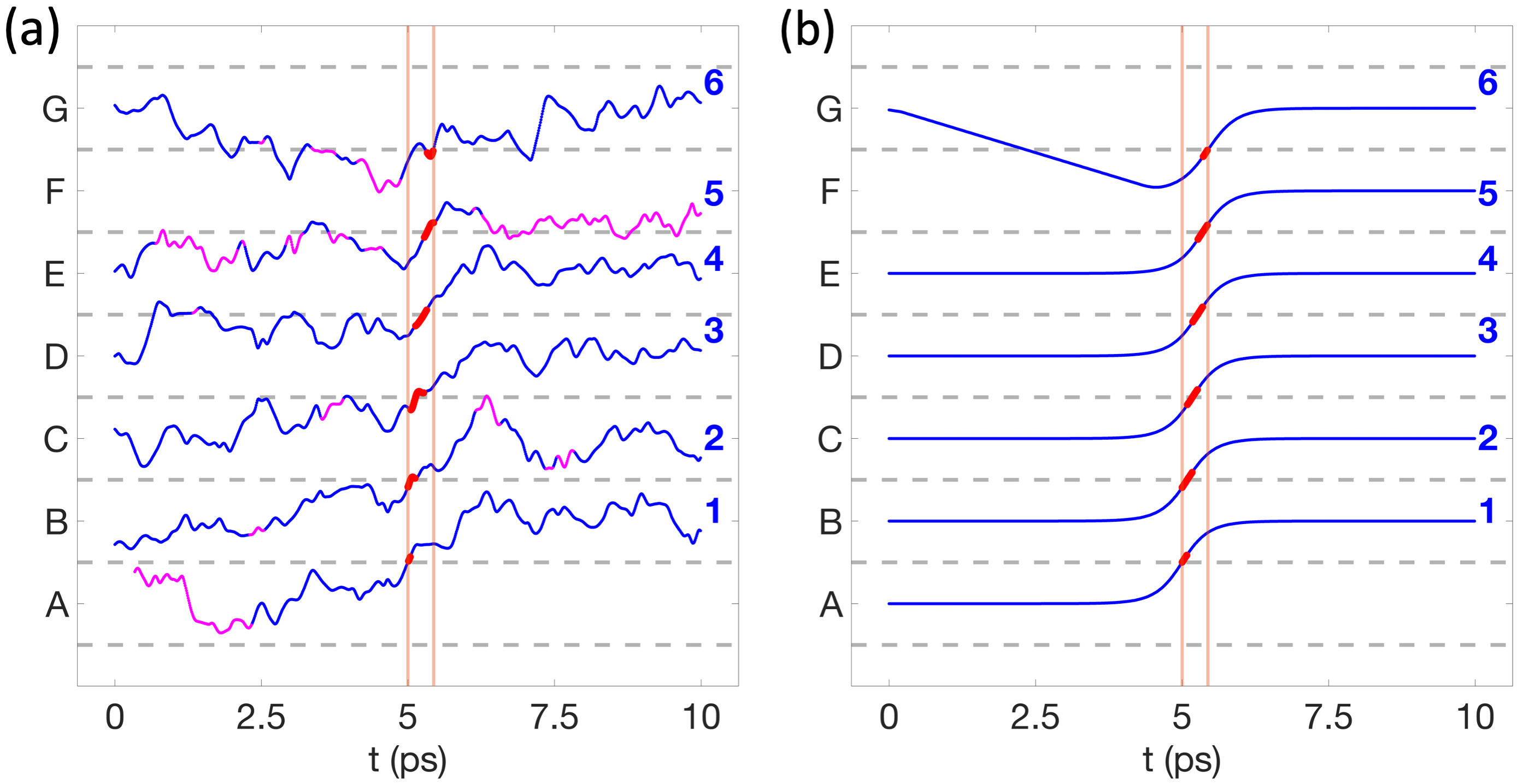}
  \centering
   \caption{Atomic trajectories illustrating coordinated atomic motion into or out of doubly-occupied cells. Panel (a) is an excerpt from our MD data, simplified in (b) to highlight the advance of each liquid atom 1-5 from one equilibrium site to the downstream site. Each atom 1-5 is advanced one substrate lattice spacing in the $x$-direction by the passage of the doubly-occupied cell, instantaneously located at the thickened red portion of the atomic trajectories. The thin vertical orange lines mark the first and final time of the occurrence of the doubly-occupied cell. 
    The $x$-boundaries between cells are marked by dashed gray lines.
    Note that the liquid atoms tend to remain near the substrate minima located halfway between adjacent dashed lines. Magenta portions of the atomic trajectories mark when the atom has drifted out of the first liquid layer. Here, the wall speed is $U_{\mathrm{WALL}}=20$ m/s, $r_{0}^{\mathrm{LL}}/\lambda=1$, and the atomic positions are sampled at every 10 fs.}
\label{fig:X_Particle_Trajectories_Kink}
\end{figure}

To highlight the sequential nature of the atomic motion involving doubly-occupied cells, we show in figure \ref{fig:X_Particle_Trajectories_Kink} the $x$-positions of six neighboring liquid atoms as a function of time as the atoms hop in the $x$-direction between doubly-occupied sites.
The atomic positions shown are obtained from the same simulation that is used to generate figure~\ref{fig:Surface_Vis}.
The actual atomic trajectories are shown in panel (a); a schematic representation of the motion is shown in panel (b) for clarity. 

In this example, atom 1 starts in cell A and hops to cell B near $t=5$ ps, creating a doubly-occupied cell, which is indicated by highlighting the trajectories for atoms 1 and 2 in red. The interaction of atoms 1 and 2 results in atom 2 hopping from cell B to cell C (already occupied by atom 3). In this manner, the doubly-occupied site propagates from cell B to cell F. The sequence terminates when atom 6 moves into the unoccupied cell G at the top of the diagram. The sequence of doubly-occupied cells exists during the time interval between the two vertical orange lines, for about 0.5 ps. During this period, the doubly-occupied cell translates by five substrate wavelengths, and five atoms propagate by one lattice spacing forward. 

\section{Frenkel-Kontorova Model of Slip} \label{sec.FK_kinks}

The two atoms in doubly-occupied cells are part of a larger number of atoms whose motion is correlated. When properly visualized, this group of atoms is seen to take on a shape that is characteristic of nonlinear waves in the sine-Gordon equation and Frenkel-Kontorova (FK) model (\cite{Braun1998}).

We define an offset for each atom near the doubly-occupied cell,
\begin{equation}
\label{eq:offset def}
\text{[offset of atom $k$]}=\frac{x_k(t)-k\lambda}{\lambda},
\end{equation}
where $x_k(t)$ is the $x$-coordinate of the atom $k$, and $k\lambda$ is the position of the atom's initial substrate equilibrium site.

\begin{figure}
  \hspace*{-0.3in}
  \centering
  \includegraphics[scale=0.40]{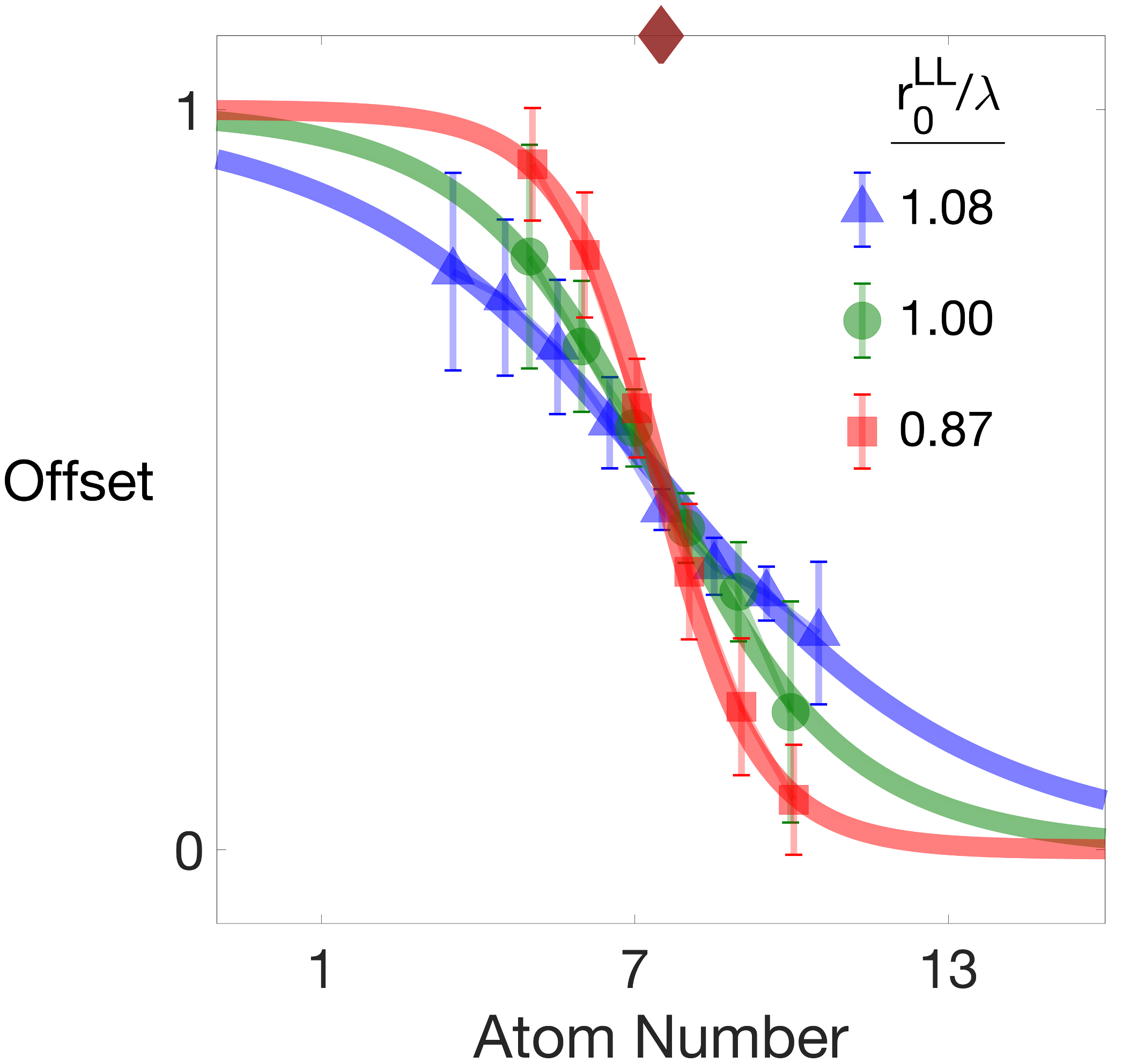}
  \centering
   \caption{Time-averaged kink profiles derived from atomic positions in the first liquid layer of the MD simulations. Data points of blue triangles, green circles, and red squares are the mean values of atomic offsets from MD simulations with $r_{0}^{\mathrm{LL}}/\lambda$  for 1.08, 1.00, and 0.87, respectively. The wave profiles are well-described by 
   (\ref{eq:kink_fitting}) shown as solid curves, with steepness $A=0.27$ (blue triangles), $A=0.44$ (green circles) and $A=0.85$ (red squares). 
   The location of the doubly-occupied cell that marks the center of the propagating kink is shown by the diamond at the top of the figure.}
\label{fig:Offset_Propagation_MD}
\end{figure}

Figure~\ref{fig:Offset_Propagation_MD} shows the offsets of the atoms in the neighborhood of a propagating doubly-occupied cell.
Downstream of the doubly-occupied cell, atoms gradually move away from their equilibrium sites as they are swept into the doubly-occupied cell.  
As the doubly-occupied cell propagates onward, upstream atoms relax into new equilibrium sites after being transported one cell forward.
These profiles, such as shown in figure \ref{fig:Offset_Propagation_MD},
thereby refer to the distribution of
atomic offsets relative to the atoms' initial equilibrium positions.
So, an offset of 0 shows that atom $k$ has not moved 
away from its initial site,
while an offset of 1 indicates that the atom has been transported forward by the distance of one substrate cell. 
Thus, the doubly-occupied cell marks a larger propagating structure of displaced atoms that we call a \textit{kink}.

The profiles seen in figure \ref{fig:Offset_Propagation_MD} can be fit to the sigmoid function,
\begin{equation}
\label{eq:kink_fitting}
\text{[offset of atom $k$]}=\frac{2}{\pi}\tan^{-1} \biggl\{ \exp\Big[-A(k-k_0)\Big]\biggl\},
\end{equation}
where the parameters $A$ and $k_0$ describe the steepness of the profile and the location of its center, respectively. This function is the discrete version of the soliton solution to the sine-Gordon equation, which is the continuum limit of the standard FK model (\cite{Braun1998}).

For each of the three curves shown in figure \ref{fig:Offset_Propagation_MD}, the data points are 
the time-averaged atomic positions over the lifetime of one kink. 
All frames are centered by aligning the location of the doubly-occupied cell. 
The curves are then generated by fitting the averaged atomic positions to the sigmoidal (\ref{eq:kink_fitting}) to find $A$ and $k_0$ for each profile.

To understand the shape and behavior of the kinks observed in the MD simulations, we model the first liquid layer using a modified version of the one-dimensional FK model. 
The FK model describes $N$ particles along the $x$-axis at positions $x_k(t)$, $k=1, 2, \ldots{} N$
that are interacting through a linear spring force with their neighboring particles and are also interacting with a sinusoidal force that arises from the periodic potential of the adjacent substrate (\cite{Braun1998}). 
If the spacing between the particles is precisely equal
to the wavelength of the potential, then the particles achieve their lowest-energy state with each particle at the periodically-located potential minima.
When an additional particle is present, such that there is one too few minima to host each of the particles, the FK model predicts that 
the spacing between particles is no longer uniform.
While most particles remain periodically placed close to substrate potential minima,
a few particles within a narrow spatial interval are crowded together. 
Thus, the FK model can be used to understand the profiles shown in figure \ref{fig:Offset_Propagation_MD}.

For our application, the particles of the FK model are the liquid atoms and the potential is due to the substrate of solid atoms.
Prior work showed that the FK model provides a description 
of the dependence of slip on shear rate (\cite{Lichter2004,Martini2008b}).
Furthermore, predictions of the FK model agree with the observations that slip is 
bounded in the limit of large shear rates over thermalized 
solid substrates (\cite{Martini2008a}).
Though these studies gave evidence that the FK model could describe certain aspects of mean slip, there was little comparison with the observations at the molecular scale.
This was due to the lack of data on molecular trajectories and the absence of a conceptual framework that permitted the localized propagating structures to be identified and extracted from the noisy MD data.
These deficiencies have been overcome in the present work through the recognition of the doubly-occupied cell.

To more realistically model the interactions between neighboring atoms in a liquid, we replace the linear nearest-neighbor coupling of the classical FK equation with a nearest-neighbor force arising from the 12-6 Lennard-Jones potential.
In addition, the liquid atoms are forced into motion by a shear force of the form
$\eta_{\mathrm{LL}}(V-\dot{x}_k)$, where $V$ is the constant velocity of the liquid layer above and $\eta_{\mathrm{LL}}$ is the friction coefficient between the adjacent liquid layers.  
As the liquid atoms move with speed $\dot{x}_k$ over the substrate, they are also subject to a frictional force 
proportional to their speed $\eta_{\mathrm{LS}}\,\dot{x}$, where $\eta_{\mathrm{LS}}$ is the liquid-solid friction coefficient. 

With these modifications, the FK model becomes
\begin{multline} \label{eq:FK_nondim}
    \ddot{x}_k=-\sin{(x_k)}\\
    + g\left[\left(\frac{\sigma}{x_{k+1}-x_k}\right)^7-
    2\left(\frac{\sigma}{x_{k+1}-x_k}\right)^{13}
-\left(\frac{\sigma}{x_k-x_{k-1}}\right)^7+
    2\left(\frac{\sigma}{x_k-x_{k-1}}\right)^{13}
    \right]\\
    + \eta_{\mathrm{LL}}(V-\dot{x}_k)
    - \eta_{\mathrm{LS}}\,\dot{x}_k ,
\end{multline}
where
\begin{equation} \label{eq:FK_parameters_nondim}
g=\frac{12\,\epsilon_{\mathrm{LL}}\lambda}{\pi \sigma_{\mathrm{LL}} \phi_{0}} \quad \mathrm{and} \quad \sigma = \frac{2\pi\sigma_{\mathrm{LL}}}{\lambda},
\end{equation}
and where the particle positions and time have been nondimensionalized in the usual way (\cite{Braun1998}).
The non-dimensional parameter $g$ signifies the strength of the liquid-liquid interaction relative to that of the liquid-substrate interaction.
The ratio of the characteristic size of the liquid atoms relative to the lattice spacing, 
$\sigma$, plays an important role in setting the ground state and the dynamics of the solutions to the FK equation (\cite{Aubry1983,Braun1998}).

The dimensionless parameters defined in (\ref{eq:FK_parameters_nondim})  depend on the dimensional liquid-liquid LJ energy, $\epsilon_{\mathrm{LL}}$, the length parameter of the liquid-liquid interaction, $\sigma_{\mathrm{LL}}$, the substrate wavelength $\lambda$, and the dimensional amplitude of the sinusoidal substrate potential energy, $\phi_{0}$. While the first three can be taken directly from the input parameters for the MD simulations, the substrate potential amplitude used in the FK model must be estimated from the MD interaction potential between all the substrate atoms and an atom in the first liquid layer. We do this by calculating the potential energy above two 
$(x,y)$-locations in the MD simulations: (i) a substrate equilibrium site, yielding 
$\phi_{\mathrm{eqb}}(z)$, and (ii) halfway between two neighboring equilibrium sites, yielding 
$\phi_{\mathrm{jump}}(z)$. 
Atoms in the first liquid layer of the MD simulations are expected to tend towards heights at which the potentials are at a minimum. We therefore take the substrate potential amplitude for the modified FK model to be one-half of the difference between the minimum values of $\phi_{\mathrm{eqb}}(z)$ and $\phi_{\mathrm{jump}}(z)$, yielding
\begin{equation}
    \phi_{0} = \frac{
    \min{\left[ \phi_{\mathrm{jump}}(z)\right]} - 
\min{\left[\phi_{\mathrm{eqb}}(z) \right]}}{2} \, .
\end{equation}
This results in a substrate potential amplitude for the modified FK model of $\phi_{0}/k_{\mathrm{B}} \simeq 35.5(1.1-r_{0}^{\mathrm{LL}}/\lambda) + 66 $ K.

\begin{figure}
  \hspace*{-0.3in}
  \centering
  \includegraphics[scale=0.38]{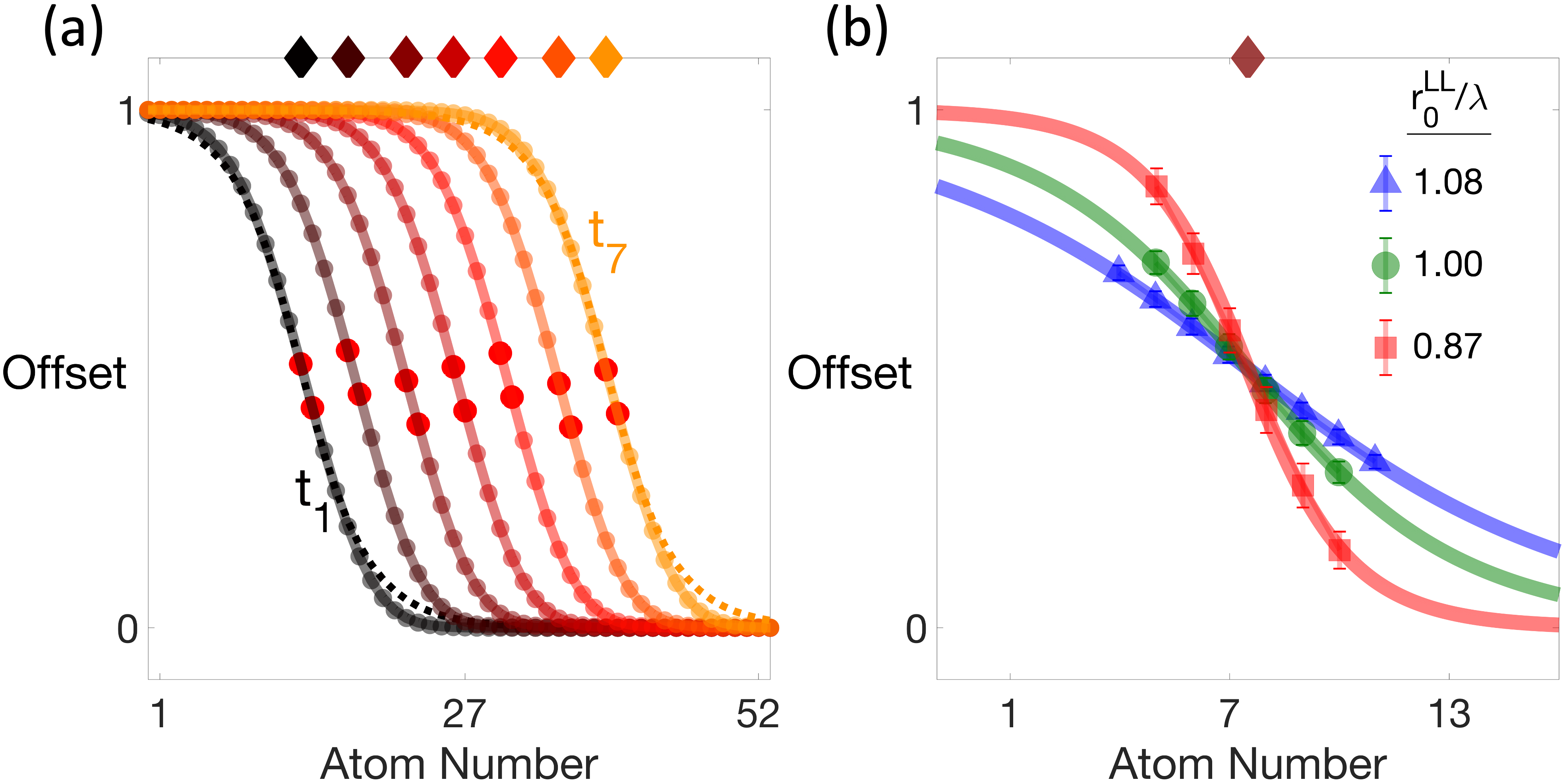}
  \centering
   \caption{Numerical results from the FK model.
   (a) The propagation of a single FK kink through 52 contiguous atoms is shown at intervals of 0.7 ps. The two atoms instantaneously inside a doubly-occupied cell are highlighted in red and the location of their center of mass is shown by a diamond at the top of the figure. 
   The dotted lines at the initial and final frames are the 
   nearly identical sine-Gordon soliton profiles, fit with $A=0.27$ for $r_{0}^{\mathrm{LL}}/\lambda=1$.
   (b) Time-averaged FK kink profiles for three different values of $r_{0}^{\mathrm{LL}}/\lambda$ showing the steepening profile as $r_{0}^{\mathrm{LL}}/\lambda$ decreases. Blue triangles: $r_{0}^{\mathrm{LL}}/\lambda = 1.08, A=0.17$; Green circles: $r_{0}^{\mathrm{LL}}/\lambda = 1.00, A=0.27$; Red squares: $r_{0}^{\mathrm{LL}}/\lambda = 0.87, A=0.56$. Data points are from numerical solutions to the FK model and correspondingly colored curves are best fits from the sigmoid function (\ref{eq:kink_fitting}).
   }
\label{fig:Offset_Propagation_FK}
\end{figure}

We use the modified FK model to describe the dynamics of the atoms near a doubly-occupied substrate cell. As in our MD simulations, we use periodic boundary conditions in the modified FK model, with a periodic domain length of $(N-1)\lambda$, where $N$ is the number of particles. This assures that there is an extra particle in the FK chain relative to the number of substrate potential minima. Figure~\ref{fig:Offset_Propagation_FK}(a) shows numerical solutions of the FK equations (\ref{eq:FK_nondim}) at multiple instants of time.
We call these nonlinear wave solutions to the FK model as 
\textit{FK kinks}.
FK kinks propagate at a constant speed with an invariant profile.
The FK kink profiles are closely approximated by the sine-Gordon soliton solution (dotted curves) from the sigmoid function (\ref{eq:kink_fitting}).
The profile can be succinctly characterized by its steepness parameter $A$, which indicates the narrowness of the FK kink profiles, 
three of which are shown in 
figure~\ref{fig:Offset_Propagation_FK}(b). 

\begin{figure}
  \hspace*{-0.20in}
  \centering
  \includegraphics[scale=0.45]{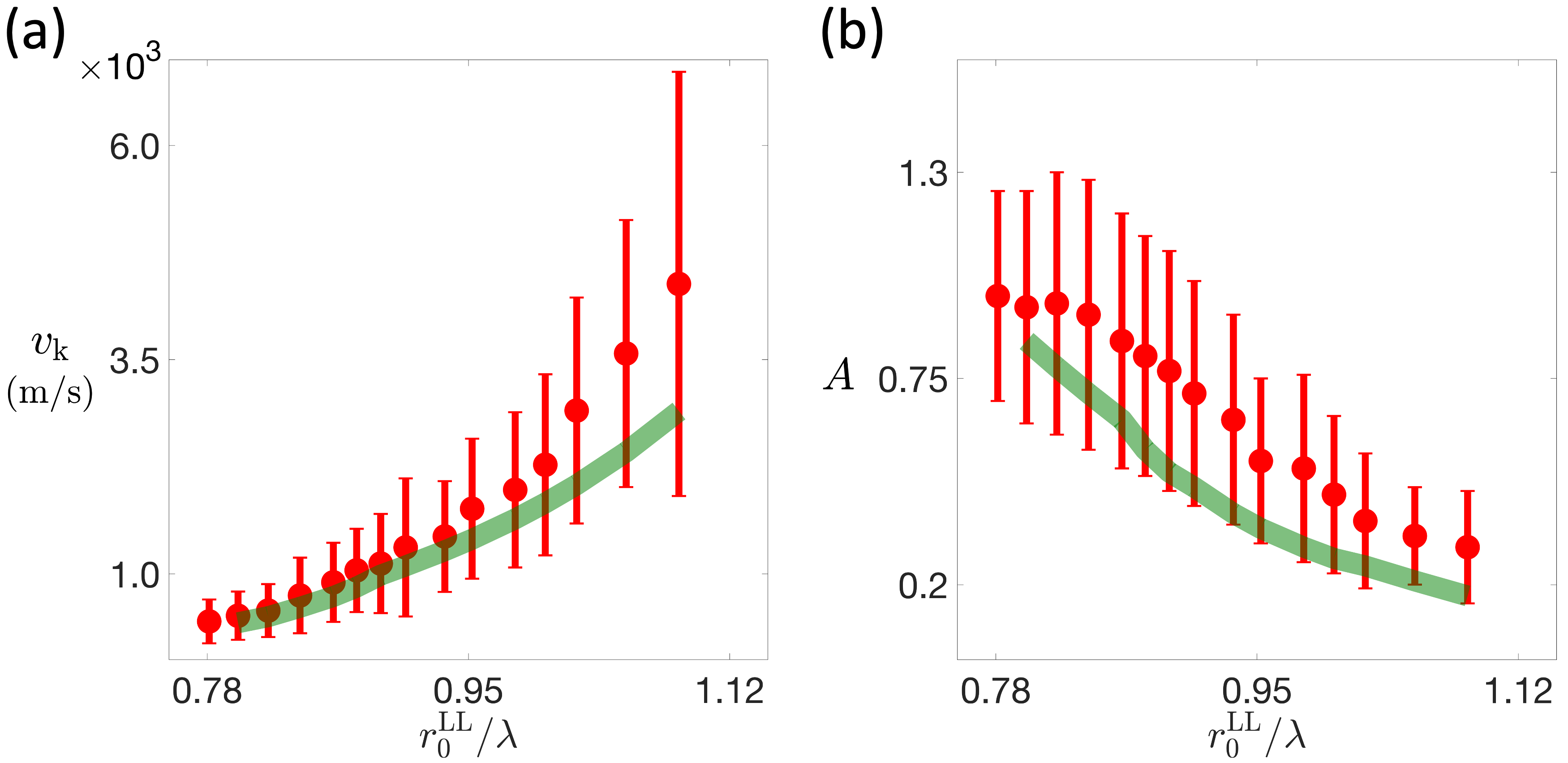}
  \centering
   \caption{Results from the MD simulations compared with those from the FK model. (a) The kink velocity, $v_{\mathrm{k}}$, as a function of $r_{0}^{\mathrm{LL}}/\lambda$. Filled red circles indicate the average MD kink velocities. The thick green line indicates the velocities of the FK kinks obtained from the FK simulations. (b) The steepness parameter, $A$, of the kinks as a function of $r_{0}^{\mathrm{LL}}/\lambda$. 
   MD data (filled red circles) was averaged over all kinks that propagated at least four substrate wavelengths. Error bars are the standard deviations of all those kinks for each $r_{0}^{\mathrm{LL}}/\lambda$.
   For the green line of the FK model, 
we set $V=1$, and use values 
$\epsilon_{\mathrm{LL}}/\mathrm{k}_{\mathrm{B}}=188\,
\mathrm{K}$, 
$\sigma_{\mathrm{LL}}=(2\pi)2^{-1/6}$, 
$\phi_{0}/\mathrm{k}_{\mathrm{B}}\simeq70\,\mathrm{K}$
identical to those of the MD simulation.
The friction factors were chosen to comply with the high shear rate limit of the FK equation,
$\eta_{\mathrm{LL}}/\eta_{\mathrm{LS}}={\cal O}(1)$, see the Appendix.
}
\label{fig:Kink_Velocity_Slope_JFM}
\end{figure}

To compare the wave properties of MD kinks to FK kinks, we compare their velocities and narrowness.
Figure \ref{fig:Kink_Velocity_Slope_JFM}(a) shows the velocity of the kinks, $v_{\mathrm{k}}$, in both the MD and FK simulations as a function of $r_{0}^{\mathrm{LL}}/\lambda$. 
We calculate kink velocities in the MD simulations by dividing the kink displacements in the $x$-direction with their lifetimes. 
It is observed that the kink velocities in MD simulations increases as $r_{0}^{\mathrm{LL}}/\lambda$ increases. 
This trend is also observed in our FK simulations. 
The agreement between the kink velocities from MD and FK simulations gets better as $r_{0}^{\mathrm{LL}}/\lambda$ decreases, where we can sample a larger number of propagating kinks, which is signified by smaller error bars. 
Figure~\ref{fig:Kink_Velocity_Slope_JFM}(b) shows the steepness parameter $A$ of the kinks of both the MD and FK simulations as a function of $r_{0}^{\mathrm{LL}}/\lambda$. 
Kinks get narrower as $r_{0}^{\mathrm{LL}}/\lambda$ decreases. 

When compared with the predictions of the modified FK model, both the profile shape, as measured by $A$, and the velocity of the kinks seen in the MD simulations, show similar trends and comparable values as a function of $r_{0}^{\mathrm{LL}}/\lambda$.
In assessing the agreement between the two sets of data, the extreme simplicity of the FK model should be recalled.
In the FK model, particles move only in one dimension and the system is closed.
This is to be contrasted with the three-dimensional motion of atoms in the MD simulations in which atoms continually move into and out of the first liquid layer.
In the FK model, the particles experience the substrate through a potential with a simple sinusoidal variation.
This is to be compared with a substrate potential due to the summation of the Lennard-Jones potentials of the solid substrate atoms. 
Not only is the Lennard-Jones potential itself more complex than that of the sinusoidal FK potential, but the potential experienced by the liquid atoms further varies as they fluctuate along their $y$- and $z$-degrees of freedom.
Additionally, the MD simulation incorporates finite temperature and heat fluxes that are absent in the FK model.
Given these differences, the rather striking degree of similarity 
delivered by the FK model can be taken as evidence of the importance of geometry, particularly the ratio $r_{0}^{\mathrm{LL}}/\lambda$, in governing the dynamics.

We lastly note that unlike the coordinated kink motion described above, atomic motion into an unoccupied 
cell does not create the same extended atomic profile as seen in figure \ref{fig:Offset_Propagation_FK} and is not described by an FK kink. 
This difference is due to the asymmetry of the Lennard-Jones potential in the modified FK model, (\ref{eq:FK_nondim}). 
The Lennard-Jones potential produces the nonlinear wave profile of a kink because its nonlinear repulsive force is huge for liquid atoms brought close together. 
In contrast, it produces no such profile in the neighborhood of a vacant cell, because it is only slightly attractive for liquid atoms separated across a vacant cell.  
 
\section{Conclusions} \label{sec.conclusions}

When sheared, liquid atoms move across a solid substrate in a motion known as 
slip.
Slip has often been assumed to be diffusive, as liquid atoms move from
one low-energy substrate site to another, independent of each other. 
However, there are conditions when slip takes place differently.
Groups of atoms form into localized nonlinear waves that propagate at great 
speeds over the substrate---orders of magnitude faster than the slip velocity. 

In this work, we focused on the atomic-level mechanisms by which liquid slip 
occurs, by systematically varying substrate lattice spacing $\lambda$ and 
characterizing the mechanisms of slip as a function of the parameter 
$r_{0}^{\mathrm{LL}}/\lambda$. 
We have shown that there is a slip velocity minimum at 
$r_{0}^{\mathrm{LL}}/\lambda=1$.
Furthermore, under certain conditions, such as for 
$r_{0}^{\mathrm{LL}}/\lambda<1$, 
slip occurs predominantly due to wave propagation.
That is, the mass transport due to slip is predominantly conveyed by 
localized nonlinear waves, and not through surface diffusion.
We have further shown that the observed waves are well-described 
in their speed and profile by solutions to a modified Frenkel-Kontorova model and its 
continuum approximation, the sine-Gordon equation. 
Conversely, when $r_{0}^{\mathrm{LL}}/\lambda>1$, liquid slip predominantly 
occurs due to mass propagation by individual atoms moving into unoccupied 
substrate sites, that is, through surface diffusion. 
Slip under all circumstances is due to the sum of the contributions from these 
waves and the isolated motions of atoms into unoccupied substrate sites. 

We have developed a novel conceptual framework 
to arrive at these conclusions.
Using the doubly-occupied cell as a numerical marker, we 
could identify individual localized propagating kinks in the noisy MD data.
Fast 10 fs sampling was used to track kinks, revealing their well-defined
profiles and fast speeds, that are comparable to predictions from 
the Frenkel-Kontorova equation modified to use Lennard-Jones interactions 
between the liquid atoms. 

Solid-on-solid friction has been modeled using two 
opposing periodic lattices.
As one lattice is forced over the other, strain deforms
the lattices leading to a 
misfit between the wavelengths of the two lattices that
propagates as a kink (\cite{Braun2001,Woods2014}).
\cite{Sisan2014} considered a nanotube 
composed of hexagonal rings of carbon atoms containing a single file of TIP3P
water molecules.
With an applied pressure gradient, episodic pulses of flow occurred due to
Frenkel-Kontorova solitons that propagated along the length of the nanotube.
These very different scenarios share with the present
work two unequal length scales
whose competition yields propagating kinks.
These related researches with different materials and with
various substrate potentials
suggest that Frenkel-Kontorova type kinks may
contribute to slip in more general settings than the 
cubic lattice structure studied here.
Many more studies and analyses will be required to gauge the
robustness of kink propagation under conditions other 
than the singular choices presented here.

Perhaps there are practical applications of the new 
understanding of the slip presented here.
Patterning the substrate to promote kinks may 
enhance slip or facilitate mixing at small scales.
At a moving contact line, the inherent nonuniformity of relative atomic 
positions and speed of motion might be a meaningful 
missing contribution in treating the microscopic origins of contact line 
motion, microbubbles and 
the nucleation of cavitation.
The inhomogeneous distribution of momentum transfer at 
the small scales investigated here may also produce 
variations of heat transfer rate at the surface that could be usefully 
exploited.
As a final speculation, we note that liquid-solid interfaces are home to 
commercially important catalysis.
The kinetics of some reactions are found to confound typical descriptions in 
terms of surface diffusion.
Furthermore, the catalytic dynamics on substrates that are densely populated by 
reactants and products are poorly understood.
Perhaps the new understanding of atomic motion at the liquid-solid interface 
will provide a useful point of view to help explain the complex cooperative 
events observed during catalysis on surfaces (\cite{Henss2019}).
\vspace{0.15in}

\section*{Appendix}
For the results of the FK model shown by the green lines in figure \ref{fig:Kink_Velocity_Slope_JFM}, 
we set $V=1$, $\eta_{\mathrm{LL}}=0.30$, $\eta_{\mathrm{LS}}=0.25$, and use values 
$\epsilon_{\mathrm{LL}}/\mathrm{k}_{\mathrm{B}}=188\,
\mathrm{K}$, 
$\sigma=(2\pi)2^{-1/6}$, and
$\phi_{0}/\mathrm{k}_{\mathrm{B}}\simeq70\,\mathrm{K}$,
identical to those of the MD simulation.
The nondimensional friction coefficients, 
$\eta_\mathrm{LS}$ and $\eta_\mathrm{LL}$, were chosen to comply with the high shear rate limit of 
the FK equation,
$\eta_{\mathrm{LL}}/\eta_{\mathrm{LS}}={\cal O}(1)$, as follows.

The friction coefficients of the FK equation are related to, but not equivalent, to the liquid viscosity. 

The equations for the slip length,
\begin{equation}
b = \frac{U_{\mathrm{FLL}}}{\dot{\gamma}},
\end{equation}
the dimensional liquid-solid friction coefficient, 
\begin{equation}
f_{\mathrm{LS}} = \frac{\tau_{\mathrm{LS}}}{U_{\mathrm{FLL}}},
\end{equation}
where $\tau_{\mathrm{LS}}$ is the shear stress measured
at the wall,
and the shear rate,
\begin{equation}
\dot{\gamma} = \frac{\tau_{\mathrm{LS}}}{\mu},
\end{equation}
where $\mu$ is the bulk viscosity,
can be combined as,
\begin{equation}\label{eq:b}
f_{\mathrm{LS}} = \frac{\mu}{b}.
\end{equation}

In a similar manner, combining the three equations, 
\begin{eqnarray}
\dot\gamma = \frac{U_{\mathrm{SLL}}-U_{\mathrm{FLL}}}{d},\\
\dot\gamma = \frac{\tau_{\mathrm{LL}}}{\mu},\\
\tau_{\mathrm{LL}} = f_{\mathrm{LL}}(U_{\mathrm{SLL}}-U_{\mathrm{FLL}}),
\end{eqnarray}
where $d$ is the atomic spacing between the first and second liquid layers, and $U_{\mathrm{SLL}}$ is the dimensional velocity of the second liquid layer, yields,
\begin{equation} \label{eq:d}
f_{\mathrm{LL}} = \frac{\mu}{d}.
\end{equation}
Hence, from (\ref{eq:b}) and (\ref{eq:d}),
\begin{equation}
	\frac{f_{\mathrm{LS}}}{f_{\mathrm{LL}}} = 
	\frac{\eta_{\mathrm{LS}}}{\eta_{\mathrm{LL}}} =
	\frac{d}{b} = {\cal O}(1).
\end{equation}

Furthermore, in the high shear rate limit of the 
nondimensional FK equation,
\begin{equation}
\dot{x} = \frac{\eta_{\mathrm{LL}}}{\eta_{\mathrm{LL}}+\eta_{\mathrm{LS}}}V.
\end{equation}
Hence, for a given shear rate, the soliton 
speed depends only on the ratio
$\eta_{\mathrm{LL}}/\eta_{\mathrm{LS}}={\cal O}(1)$. 

Finally, using $\mu =1.96\times10^{-4}$~kg/m-s
(\cite{meier2004transport}),
$d = 3\times 10^{-10}$~m,
$\eta_{\mathrm{LL}}/\eta_{\mathrm{LS}}={\cal O}(1)$,
and nondimensionalizing  
$\eta_{LS} = \mu \,d\,\lambda/(2\pi\sqrt{\phi_0 \,m})$,
\begin{equation}
    \eta_{LS} \approx 0.74.
\end{equation}

\noindent 
\textbf{Acknowledgements.} This research was supported in part through the computational
resources and staff contributions provided for the Quest
high-performance computing facility at Northwestern University
which is jointly supported by the Office of the Provost, the
Office for Research, and Northwestern University Information
Technology.\\ 

\noindent 
\textbf{Funding.} This research was supported by a Fellowship from the 
McCormick School of Engineering and the Department of Mechanical Engineering, 
Northwestern University. \\

\noindent 
\textbf{Declaration of interests.} 
The authors report no conflict of interest.\\

\noindent 
\textbf{Author ORCIDs.}\\
Metehan \c{C}am https://orcid.org/0000-0001-7362-3063\\
Christopher G. Goedde https://orcid.org/0000-0001-5493-3584\\
Seth Lichter https://orcid.org/0000-0001-9124-9976

\clearpage

\bibliographystyle{jfm}
\bibliography{JFMBIB}
\clearpage
\noindent 

\end{document}